\documentclass[12 pt]{article}
\usepackage{hyperref}
\usepackage{url}
\usepackage{graphicx,subfigure,epsfig,psfrag}
\usepackage{amsmath,amssymb,bm,bbm}
\usepackage{verbatim,float,url,enumerate}
\usepackage{graphicx,subfigure,epsfig,psfrag}
\usepackage{harvard}
\usepackage[ruled]{algorithm2e}
\usepackage{algpseudocode}
\usepackage{booktabs}
\usepackage{array}
\usepackage{makecell}
\newcolumntype{x}[1]{>{\centering\arraybackslash}p{#1}}
\usepackage{tikz}
\newcommand\diag[4]{%
  \multicolumn{1}{p{#2}|}{\hskip-\tabcolsep
  $\vcenter{\begin{tikzpicture}[baseline=0,anchor=south west,inner sep=#1]
  \path[use as bounding box] (0,0) rectangle (#2+2\tabcolsep,\baselineskip);
  \node[minimum width={#2+2\tabcolsep-\pgflinewidth},
        minimum  height=\baselineskip+\extrarowheight-\pgflinewidth] (box) {};
  \draw[line cap=round] (box.north west) -- (box.south east);
  \node[anchor=south west] at (box.south west) {#3};
  \node[anchor=north east] at (box.north east) {#4};
 \end{tikzpicture}}$\hskip-\tabcolsep}}

\newcommand{\be}{\begin{equation}}
\newcommand{\ee}{\end{equation}}
\newcommand{\mR}{\mathbbm{R}}

\newcommand{\bI}{\mathbf{I}}
\newcommand{\bX}{\mathbf{X}}
\newcommand{\bu}{\mathbf{u}}

\newcommand{\bW}{\mathbf{W}}
\newcommand{\by}{\mathbf{y}}
\newcommand{\bp}{\mathbf{p}}

\newcommand{\bz}{\mathbf{z}}
\DeclareMathOperator*{\argmin}{argmin\,}
\newcommand{\bx}{\mathbf{x}}
\newcommand{\bZ}{\mathbf{Z}}
\newcommand{\btheta}{\boldsymbol{\theta}}
\newcommand{\bmu}{\boldsymbol{\mu}}
\newcommand{\bbeta}{\boldsymbol{\beta}}
\newcommand{\bprox}{\mathbf{prox}}
\newcommand{\bobeta}{\boldsymbol{\beta}_{\text{old}}}

\begin{document}

\title{An Ordered Lasso and Sparse Time-lagged Regression}
\author{Xiaotong Suo\thanks{email:xiaotong@stanford.edu} \and Robert
 Tibshirani\thanks{email:tibs@stanford.edu, Supported by NSF Grant DMS-99-71405 and National Institutes of Health Contract
N01-HV-28183}}
\date{\normalsize Institute for Computational
\& Mathematical Engineering,\\ and Departments of Health Research \& Policy, and Statistics, Stanford University  }
\maketitle
\begin{abstract}
We consider a regression scenario where it is natural to impose  an order constraint
on the coefficients. We propose an order-constrained version of 
$\ell_1$-regularized regression (lasso)  for this problem, and show how to solve it efficiently using the well-known
Pool Adjacent Violators Algorithm as its proximal operator. The main application of this idea
is to  time-lagged regression, where we predict an outcome at time $t$ from features
at the previous $K$ time points. In this setting it is natural to assume that the
coefficients decay as we move farther away from $t$, and hence the order constraint is reasonable.
Potential application areas include financial time series and prediction
of dynamic patient outcomes based on  clinical measurements.
We illustrate this idea on real and simulated data.
\end{abstract}
\section{Introduction}
\label{sec:introduction}
Suppose that we observe data $(\bx_i, y_i)$ for $i=1,2, \ldots ,N$
where $N$ is the number of observations, $\bx_i=(x_{i1},x_{i2},\ldots, x_{ip})$ is a vector of $p$ feature measurements, and
$y_i$ is a response value. We consider the usual linear regression framework
\begin{eqnarray*}
y_i= \beta_0 +\sum_{j = 1}^p x_{ij}\beta_j +\epsilon_i
\end{eqnarray*} with ${\rm E} (\epsilon_i)=0$ and ${\rm Var}(\epsilon_i)=\sigma^2$.
The lasso or $\ell_1$-regularized regression \cite{Ti96}  chooses the parameters $\beta_0, \bbeta=(\beta_1, \beta_2, \ldots \beta_p)$ to solve
\begin{eqnarray*}
{\rm minimize}\Bigl\{ \frac{1}{2}\sum_{i=1}^N\left(y_i-\beta_0-\sum_{j=1}^p x_{ij}\beta_j\right)^2+
\lambda\sum_{j=1}^p|\beta_j|\Bigr\},
\label{eq:lasso}
\end{eqnarray*}
where $\lambda\geq  0 $ is a fixed tuning parameter. This problem is convex
and yields sparse solutions for sufficiently large values of $\lambda$.

In this paper we add an additional order constraint on the coefficients, and we call
the resulting procedure the {\em ordered lasso}.
 We derive an efficient algorithm for solving the resulting problem. The main application of this idea is  to time-lagged regression, where we predict an outcome at time $t$ from features at the previous $K$ time points. In this case, it is natural to
assume that the
coefficients decay as we move farther away from $t$ so that the order (monotonicity)
constraint is reasonable. A key feature of our procedure  is that it {\em automatically} determines
the most suitable value of $K$ for each predictor, directly from the monotonicity constraint.
 The paper is organized as follows. Section \ref{sec:ordlasso} contains  motivations and algorithms for solving the ordered lasso, as well as results  comparing the ordered
and standard lasso on simulated data.  Section \ref{sec:timelag} contains the detailed
algorithms for applying the ordered lasso to the time-lagged regression. We demonstrate the usage of such algorithms on real and simulated data in Sections \ref{sec:sim} and \ref{sec:ozone}.
 We also apply this framework to  auto-regressive (AR) time series and  compare its performance
to both the traditional method for fitting AR model  using least squares and the
Akaike information criterion, and the lasso procedure for fitting AR\ model. Section \ref{sec:df}
 gives a definition of  degrees of freedom for the ordered lasso.  Section \ref{sec: logistic} generalizes
the ordered lasso to  the logistic regression model. Section \ref{sec:discussion}
contains some discussion and directions for future work. 
\section{Lasso with an order constraint}
\label{sec:ordlasso}
\subsection{The basic idea}
We consider the lasso  problem with 
an additional monotonicity constraint, i.e.,
\begin{eqnarray*}
{\rm minimize}\Bigl\{\frac{1}{2}\sum_{i=1}^N(y_i-\beta_0-\sum_{j=1}^p x_{ij}\beta_{j})^2+\lambda\sum_{j=1}^p|\beta_j|\Bigr\},
\end{eqnarray*}
subject to $|\beta_1|\geq |\beta_2|\geq \ldots\geq |\beta_p|$. This setup makes sense in problems where some natural order exists among the features. However, this problem
is not convex. Hence we modify the approach,  writing each $\beta_j$ as $\beta_j = \beta_j^+-\beta_j^-$ with 
$\beta_j^+, \beta_j^- \geq 0$. We propose the following  problem
\begin{eqnarray}
 {\rm minimize}\Bigl\{\frac{1}{2} \sum_{i=1}^N(y_i-\beta_0-\sum_{j=1}^p x_{ij}(\beta_j^+-\beta_j^-))^2+
\lambda\sum_{j=1}^p(\beta_j^+ +\beta_j^-)\Bigr\},
\label{eq:ordlasso}
\end{eqnarray}
subject to $\textstyle\beta_1^+\geq \beta_2^+ \geq \ldots \geq \beta_p^+\geq 0$ and
$\beta_1^-\geq \beta_2^-\geq \ldots \geq \beta_p^-\geq 0$.
The use of positive and negative components (rather than absolute values)
 makes this a convex problem. Its solution  typically has one or both of each pair
($\hat\beta_j^+, \hat\beta_j^-$) equal to zero, in which case $|\hat\beta_j|=
\hat\beta_j^+ + \hat\beta_j^-$ and  the solutions $|\hat\beta_j|$ are monotone non-increasing in $j$. However, this need not be the case, 
as it is possible for both $\hat\beta_j^+$ and $ \hat\beta_j^-$ to be positive
and the $|\hat\beta_j|$ to have some non-monotonicity.
In other words, the constraints strongly encourage, but don't require, that the
solutions are monotone in absolute value. A similar approach was used in the interaction  models of \citeasnoun{BTT2013}.
This problem can be solved by a standard quadratic programming algorithm, and this works
well for small problems. For larger problems, there is an efficient first-order
generalized gradient algorithm, which uses  the
Pool
Adjacent Violators Algorithm (PAVA) for isotonic regression as its proximal operator (for example, see \citeasnoun{deleeuw}).    
We describe this in the next subsection.
\subsection{Algorithmic details}
\label{subsec:orderedlassoalg}
We assume that the predictors and outcome are centered so that the intercept has
the solution $\hat\beta_0=0$. For illustrative purposes, we write our data in matrix
form. Let $\bX$ be the $N\times p$ data matrix and $\by$ be the vector of length $N$ containing the response value for each observation. We first consider the
following problem 
\begin{eqnarray}
{\rm minimize}\Bigl\{\frac{1}{2}(\by-\bX\bbeta)^T(\by-\bX\bbeta)+\lambda \sum_{j = 1}^p\beta_j\Bigr\},
\label{eqn:matrixform}
 \end{eqnarray}
subject to $\beta_1\geq \beta_2\geq\ldots \geq\beta_p\geq 0$. We let 
$h(\bbeta) = \lambda \sum_{j = 1}^p \beta_j+\mathbbm{I}_C(\bbeta)$, where $\mathbbm{I}$
 is an indicator function and $C$ is the convex set
given by $\{\bbeta\in \mR^p| \beta_1\geq \beta_2\geq\ldots \geq\beta_p\geq 0\}$.
We want to calculate the proximal mapping of $h(\bbeta)$, i.e.,
\begin{eqnarray}
\bprox_h(\bbeta) = {\argmin}_{\bu}\Bigl\{\lambda \sum_{j =
1}^pu_j+\mathbbm{I}_C(\bu)+\frac{1}{2}\|\bu-\bbeta\|^{2}\Bigr\}.
\label{eqn:proximalmapping}
 \end{eqnarray}
 There is an elegant solution to obtain this proximal mapping. We  first consider solving the following problem
 \begin{eqnarray}
{\rm minimize}\Bigl\{ \frac{1}{2}\sum_{j=1}^n (y_{i}-  \theta_i)^2+\lambda \sum_{i=1}^{n}
\theta_i\Bigr\},
\label{eq:prox}
 \end{eqnarray}
subject to $\theta_1 \geq \theta_2 \geq\ldots \geq \theta_n\geq 0$. The solution can be obtained from an  isotonic regression 
using the well-known Pool Adjacent Violators Algorithm \cite{BBBB1972}.
In particular, if $\{\hat \theta_i\}=\{\hat y_i^\lambda\}$ is the  solution to the isotonic regression
of 
$\{y_i-\lambda\}$, i.e.,
\begin{eqnarray}
\{\hat \theta_i\}={\rm argmin_{\btheta}}\Bigl\{ \frac{1}{2}\sum_{i=1}^n (y_i-\lambda-\theta_i)^2\Bigr\},
\label{eq:isotonic}
 \end{eqnarray}
 subject to $\theta_1\geq \theta_2 \geq \ldots \geq \theta_n$, then $\{\hat y_i^\lambda\cdot \mathbbm{I}(\hat y^\lambda_i>0)\}$ solves
problem (\ref{eq:prox}). 
Hence the solution to  (\ref{eqn:proximalmapping}) is
\begin{eqnarray}
\bprox_h(\bbeta) = \hat\bbeta^\lambda \cdot{\mathbbm{I}}\ (\hat{\bbeta}^\lambda >0).
\label{eq: proxoperator}
\end{eqnarray}
Using this in the proximal gradient algorithm, the first-order generalized gradient update step of $\bbeta$  for solving (\ref{eqn:matrixform}) is 
\begin{eqnarray}
\bbeta \leftarrow {\bprox}_{\gamma h}(\bbeta-\gamma \bX^T(\bX\bbeta-\by)).
\label{eq:gg}
 \end{eqnarray}

The value $\gamma>0$ is a step size that is adjusted by backtracking to ensure that
the
objective function is decreased at each step.
To solve (\ref{eq:ordlasso}) we augment each predictor $x_{ij}$
with $x^*_{ij}= - x_{ij}$ and write $x_{ij}\beta_j=x_{ij}\beta^+_j + x^*_{ij}\beta^-_j$.
We denote the expanded parameters by
$(\bbeta^+, \bbeta^-)$ and apply the proximal\ operator (\ref{eq:gg}) alternatively  to
$\bX$ and $\bX^*$
to obtain the minimizers ($\hat\bbeta^+, \hat\bbeta^-)$.
Details for solving (\ref{eq:ordlasso}) can be seen in Algorithm \ref{alg: orderedlassoalgorithm}. Isotonic regression can be computed in $O(N)$ operations \cite{grotz} and hence the
ordered lasso algorithm can be applied to large datasets. 
\begin{algorithm}
\caption{Ordered Lasso}
\label{alg: orderedlassoalgorithm}
\KwData  {$\bX\in \mR^{n\times p},\by\in \mR^n$,  \ $\bX^* = -\bX  $}
Initialize $\hat \bbeta^+, \hat \bbeta^-=0\in \mR^p, \lambda$ \;
\While{(not converged)}{
Fix ${\hat \bbeta^{-(k)}}$,
        ${\hat\bbeta^{+}}\leftarrow \bprox_{t_k\lambda}({\hat\bbeta^+}-{\hat\bbeta^-}-
        t_k\bX^T(\bX{\hat\bbeta^+}+\bX^{*}\hat\bbeta^{-}-\by));$ \
Fix ${\hat\bbeta^{+(k+1)}}$, ${\hat\bbeta^-} \leftarrow \bprox_{\tilde t_k\lambda}
    (\hat\bbeta^{+(k+1)}-\hat\bbeta^{-}-\tilde t_k\bX^{T} (\bX\hat{\bbeta}^{+(k+1)}+\bX^{*}\hat\bbeta^{-}-\by));$}
\end{algorithm}

The ordered lasso can be easily adapted to the elastic net  \cite{enet}
 and the adaptive lasso  \cite{zou2006a} by some simple modifications to the proximal
operator in Equation (\ref{eq: proxoperator}).

\subsection{Comparison between the ordered lasso and the lasso}
\begin{figure}[H]
\centering
\includegraphics[scale = 0.7]{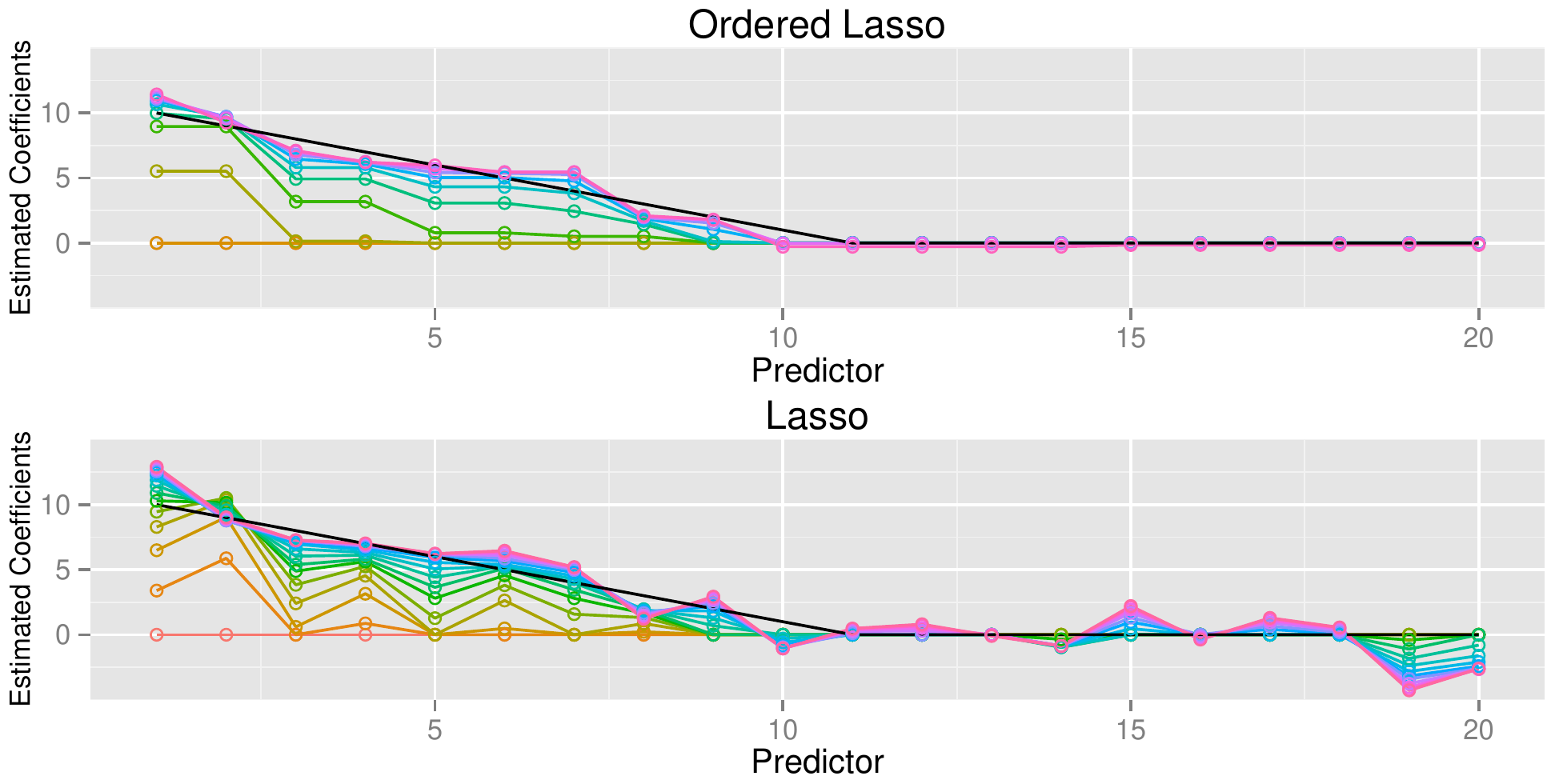}
\caption{\textit{Example of the ordered lasso compared to the standard lasso: the data was generated from  a true monotone sequence of coefficients plus Gaussian noise: $y_i= \sum_{j=1}^p x_{ij}\beta_j+\sigma\cdot Z_i$, with $x_{ij}\sim N(0,1)$,
$\bbeta=(10,9,\ldots ,2,1,0,0,\ldots 0)$, $\sigma=7$. There were 20 predictors and 30 observations. The black profiles show the true coefficients and the colored profiles are the estimated coefficients for different values of $\lambda$ from the largest(at bottom) to the smallest(at
top). }}
\label{fig:simplot}
\end{figure}

Figure \ref{fig:simplot} shows a comparison between the ordered lasso and
the standard lasso.
The data was generated from  a true monotone sequence plus Gaussian noise.
The black profiles show the true coefficients, while the colored profiles
are the estimated coefficients for different values of $\lambda$, from largest
 (at bottom) to
smallest  
(at top). The corresponding plot for the lasso is shown in the bottom panel.
The ordered lasso--- exploiting the monotonicity--- does a much better job of recovering
the true coefficients than the lasso, as seen by  the fluctuations of the estimated coefficients in the   tails of the lasso
plot.  
\subsection{Relaxation of the monotonicity requirement}
As a generalization of our approach, we can relax problem (\ref{eq:ordlasso}) as follows
\begin{eqnarray*}\textstyle
{\rm minimize} \{ \frac{1}{2}
 \sum_{i=1}^N(y_i-\beta_0-\sum_{j=1}^p x_{ij}\beta_j)^2+ \lambda\sum_{j=1}^p(\beta_j^+ +\beta_j^-)\\ \textstyle+\theta_1 \sum_{j=1}^{p-1}(\beta_{j}^+-\beta_{j+1}^+)_+ + \theta_2 \sum_{j=1}^{p-1}(\beta_{j}^--\beta_{j+1}^-)_+\},
\label{eq:ordlassog*}
\end{eqnarray*}
subject to $\beta_j^+, \beta_j^- \geq 0 , \forall j$.
As $\theta_1,\theta_2 \rightarrow \infty$, the last two penalty terms
force monotonicity and this is equivalent to (\ref{eq:ordlasso}).
However, for intermediate positive values of $\theta_1,\theta_2$, these penalties encourage near-monotonicity.
This idea was proposed in \citeasnoun{THT2011} for data sequences, generalizing the isotonic regression
problem. The authors derive an efficient algorithm {\tt NearIso} which is a generalization of the well-known
PAVA procedure mentioned above. Operationally, this creates no extra complication in our framework:
we simply use {\tt NearIso} in place of PAVA in the  generalized gradient algorithm
described  in Section \ref{subsec:orderedlassoalg}.
\label{NearIso}
\section{Sparse time-lagged regression}
\label{sec:timelag}
In this section we apply the ordered lasso to the time-lagged regression problem.
There are two problems we consider. The first one is the {\em static outcome} problem, where we observe outcome at a fixed time $t$ and predictors at a series of
 time points,
and the outcome at time $t$ is predicted from the predictors at  previous time
points. We also consider the {\em rolling prediction} problem
where we observe both outcome and predictors at a series of time points and  the outcome is predicted at each time point from the predictors at previous time points. Again, we assume that the predictors and outcome are centered so that the intercept has
the solution $\hat\beta_0=0$. Henceforth we will continue to omit the intercept.
\subsection{Static prediction from time-lagged features}
\label{subsec:statistimelag}
Here we consider the problem of predicting an outcome at a fixed time point from a set of time-lagged predictors. We assume that our data has the form
$ \{y_i, x_{i11},\ldots x_{iK1}, x_{i12},\ldots x_{iK2}, \ldots x_{i1p},\ldots ,x_{iKp}\}$, for
$i = 1,2,\ldots, N$ and $N$ being the number of observations. The value  $x_{ikj}$ is the measurement
of predictor $j$ of  observation $i$, at time-lag $k$ from the
current time $t$. In other words, 
we predict   the outcome at  time $t$
from  $p$ predictors, each measured at $K$  time points preceding the current time $t$. Our model has the form
\begin{eqnarray*}
y_i=\beta_0+\sum_{j=1}^p \sum_{k=1}^K x_{ikj}\beta_{kj}+\epsilon_i,
\end{eqnarray*}
with ${\rm E} (\epsilon_i)=0$ and ${\rm  Var}(\epsilon_i)=\sigma^2$.
We write each $\beta_{kj} = \beta_{kj}^+ - \beta_{kj}^-$ and solve 
\begin{eqnarray}\textstyle
     {\rm minimize}{\Bigl\{ \frac{1}{2}\sum_{i=1}^N (y_i-\hat y_i)^2+\lambda \sum_{j=1}^p \sum_{k=1}^{K} (\beta_{kj}^+ + \beta_{kj}^-)\Bigr\}},
\label{eq:timelagstaticobj}
\end{eqnarray}
subject to $\beta_{1j}^+ \geq \beta_{2j}^+ \geq\ldots \geq \beta_{Kj}^+\geq 0$ and $\beta_{1j}^- \geq \beta_{2j}^-\geq \ldots \geq \beta_{Kj}^-\geq 0, \forall j$.
This model makes
the plausible assumption that each predictor has an effect up to
$K$ time units away from
the current time $t$, and this effect is  monotone non-increasing as we move
farther back in time.

In order to solve (\ref{eq:timelagstaticobj}), we first write  each $\beta_{kj}^\pm$ in the following form,
\begin{eqnarray}\textstyle
\Big\{\underbrace{\beta^\pm_{11},\beta^\pm_{21},\cdots, \beta^\pm_{K1}}_{\text{block 1}}|\underbrace{\beta^\pm_{12},\beta^\pm_{22},\cdots, \beta^\pm_{K2}}_{\text{block 2}}|\cdots|\underbrace{\beta^\pm_{1p},\beta^\pm_{2p},\cdots,\beta^\pm_{Kp}}_{\text{block p}}\Big\}\nonumber.
\end{eqnarray}
This is a blockwise coordinate descent procedure, with one block for each predictor.
 For example, 
at step $j$, we compute the update for block $j$ while  holding
the rest of the blocks constant. We augment each predictor $x_{ikj}$ by $x^*_{ikj}
= -x_{ikj}$. With a sufficiently large time-lag $K$, the procedure automatically chooses an appropriate number of non-zero coefficients for each predictor,
and zeros
out the rest in each
block because of the order constraint on each predictor.  Details can be seen in Algorithm \ref{alg: statictimelaglasso}.
\begin{algorithm}
\caption{Ordered Lasso for Static Prediction}
\label{alg: statictimelaglasso}
\KwData{$\bX \in \mR^{n\times(Kp)}, \by\in \mR^{n}, \bX^*=-\bX$}
Initialize $\hat\bbeta^+,\hat\bbeta^-\in \mR^{Kp}$, $ \hat\beta^+_{kj}=0$, $\hat\beta_{kj}^- =0, \lambda$\;
\While{not converged}{
        \ForEach{$j = 1, \cdots, p$}{
               For each $i$, $r_{i} = y_i  - \sum_{\ell  \neq j}\sum_{k=1}^K (x_{i
               k\ell}\hat\beta^+_{k\ell}+x^*_{i k\ell} \hat\beta^-_{k \ell})$\;
                 Apply the ordered lasso  (Algorithm \ref{alg: orderedlassoalgorithm})\ to data  $\{r_{i},  (x_{i 1j},\ldots ,x_{i Kj}),(x_{i 1j}^*,\ldots, x_{i Kj}^*), i= 1,2,\cdots,n\}$ \\ to obtain new estimates 
                  $\{\hat\beta_{kj}, k=1,2,\ldots K\}$;}}
\end{algorithm}
\vspace*{-.4cm}
\subsection{Rolling prediction from time-lagged features}
Here we assume that our data has the form
$\{y_t, x_{t1},\ldots ,x_{tp}\}$,  for $t=1,2,\ldots , N$. In detail,  we have a  time series for which we observe
the outcome and the values of each predictor at $N$
different time points. We consider a time-lagged regression model with a maximum lag of $K$ time points
\begin{eqnarray*}
y_t=\beta_0 +\sum_{j=1}^p \sum_{k=1}^K x_{t-k,j}\beta_{kj}+\epsilon_t,
\end{eqnarray*}  
with ${\rm E} (\epsilon_t)=0$ and ${\rm Var}(\epsilon_t)=\sigma^2$.
We write each $\beta_{kj}=\beta_{kj}^+-\beta_{kj}^-$
 and  propose the following problem
\begin{eqnarray}
     {\rm minimize}\Bigl\{ \frac{1}{2}\sum_{t=1}^N (y_t-\hat y_t)^2+\lambda \sum_{j=1}^p \sum_{k=1}^{K} (\beta_{kj}^+ + \beta_{kj}^-)\Bigl\},
\label{eq:timelagorderedobj}
\end{eqnarray}
subject to $\beta_{1j}^+ \geq \beta_{2j}^+ \geq \ldots \geq \beta_{Kj}^+\geq 0$ and $\beta_{1j}^- \geq \beta_{2j}^- \geq\ldots \geq \beta_{Kj}^-\geq 0, \forall j$. To solve this problem, we convert the problem into the  form of Section \ref{subsec:statistimelag}.
We build a larger feature matrix $\bZ$ of size $N\times (K p)$,
with $K$ columns for each predictor. In detail, each row has the form
\begin{equation*}
\Big\{x_{t-1,1},x_{t-2,1}, \ldots, x_{t-K,1} | x_{t-1,2},x_{t-2,2}, \ldots, x_{t-K,2}|
\ldots |x_{t-1,p},x_{t-2,p}, \ldots, x_{t-K,p}\Big\}.
\end{equation*}
Each block corresponds to a predictor lagged
for $1,2,\ldots, K$ time units. The matrix $\bZ$ has $N$ such rows, corresponding to time points $t-1,t-2, \ldots t-K$.
Again, we augment each predictor $x_{t-k,j}$ with $x^*_{t-k,j}=-x_{t-k,j}$ and   choose a sufficiently large time-lag $K$, and let the procedure to zero out
extra coefficients for each predictor. We  can solve (\ref{eq:timelagorderedobj})  using block coordinate descent
as in the previous subsection. Details are shown in Algorithm \ref{timeLagorderedlasso}.
\begin{algorithm}
\caption{Ordered Lasso for Rolling Prediction}
\label{timeLagorderedlasso}
\KwData{$\bX \in \mR^{n\times(Kp)}$,  $\by\in \mR^{n}, \bX^*=-\bX$}
Initialize $\hat\bbeta^+,\hat\bbeta^-\in \mR^{Kp}$, $ \hat\beta^+_{kj}=0$, $\hat\beta_{kj}^-
=0, \lambda$\;
\While{not converged}{
        \ForEach{$ j = 1, \cdots, p $}{
                For each $t$, $r_{t} = y_t- \sum_{\ell \neq j}\sum_{k=1}^K (x_{t-k,\ell
}\beta^+_{k\ell}+x^*_{t-k,
\ell  } \beta^-_{k \ell})$\;
 Apply the ordered lasso (Algorithm \ref{alg: orderedlassoalgorithm}) to data $\{r_{t},(x_{t-1,j},\ldots ,x_{t-K,j}),(x_{t- 1,j}^*,\ldots
      ,x^{*}_{t- K,j}), t= 1, 2, \cdots ,n\}$ \\ to obtain new estimates $\{\hat\beta_{kj}, k=1,2,\ldots
      ,K\}$;}}
\end{algorithm}
\vspace*{-.4cm}
\subsection{Simulated example}
\label{sec:sim}
\begin{figure}[h]
\centering
\includegraphics[scale = 0.8]{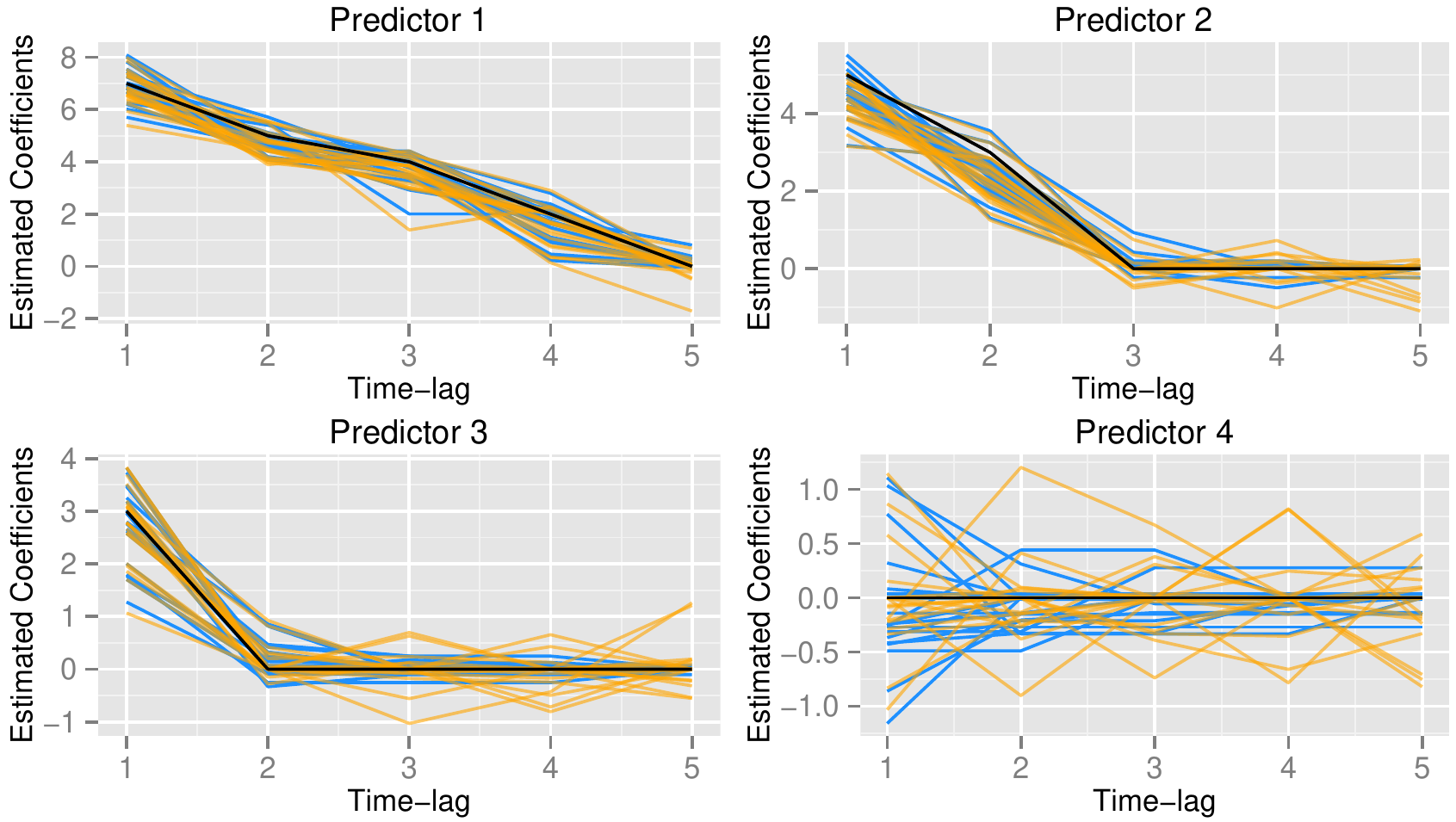}
\caption{\textit{ True coefficients (black), coefficient estimates
 the ordered lasso (blue) and the standard lasso (orange) from 20 simulations.}}
\label{fig:fig2}
\end{figure}

Figure \ref{fig:fig2} shows an example of the ordered lasso procedure
applied to a rolling  time-lagged regression. The simulated data consists of four predictors with  a maximum lag of 5 time points and 111 observations.
The true coefficients for each of the four predictors were $(7,5,4,2,0)$,
$(5,3,0,0,0)$, $(3,0,0,0,0)$ and $(0,0,0,0,0)$.
The features were generated as i.i.d. $N(0,1)$
with Gaussian noise of a   standard deviation equal to 7.
The figure shows the true coefficients (black), and estimated coefficients of the  ordered lasso (blue) and  the standard lasso (orange)  from 20 simulations. For
each method, the coefficient estimates with the smallest mean squared error (MSE)
in each realization are plotted. We see that the  ordered lasso does a better job of recovering the true
coefficients. The average  mean squared errors for the ordered lasso
and the  lasso were $4.08(.41)$ and $6.11(.54)$, respectively.

\begin{figure}
\centering
 \includegraphics[scale = 0.6]{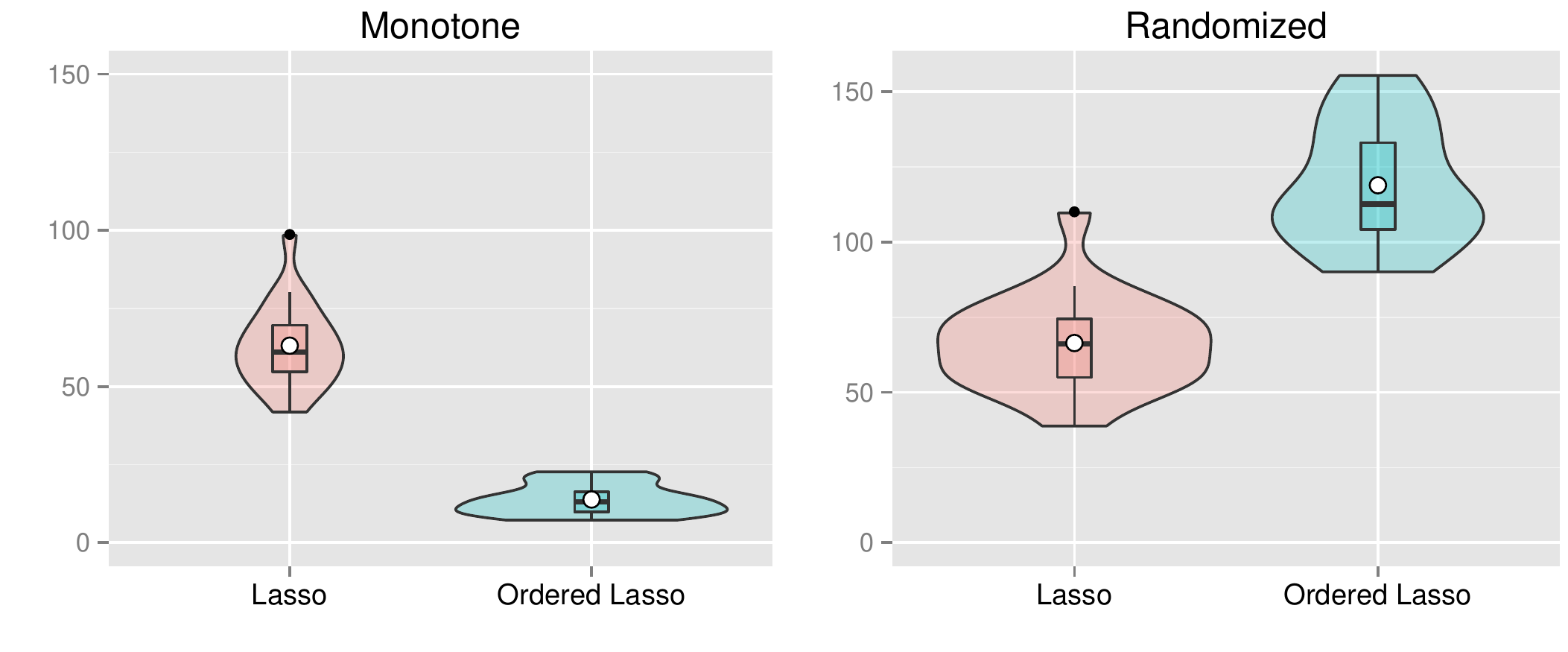}
\caption{\textit{The lasso and the  ordered lasso, applied to time-lagged features.
Shown is the mean squared error over 30 simulations using the minimizing value of $\lambda$ for
each realization. In the left panel, the true coefficients are monotone; in the right,
they have been scrambled so that monotonicity does not hold.}}
\label{fig:figbox}
\end{figure}

Figure \ref{fig:figbox} shows a larger example with a maximum lag of 20 time points.
The features were generated  as i.i.d. $N(0,1)$ with  Gaussian noise of a standard
deviation equal to 7.
Let $f(a,b,L)$ denote the equally spaced sequence from $a$ to $b$
of length $L$. The true coefficients  for each of the four predictors were 
$\{f(5,1,20)\}$, $\{f(5,1,10),f(0,0, 10)\}$,  $ \{f(5,1,5),f(0,0,15)\}$
and $\{f(0,0,20)\}$.
The left panel of the figure shows the mean squared error  of the  standard lasso
and the ordered lasso, over 30 simulations. The value of $\lambda$ giving the minimum
MSE was chosen in each realization. 
In the right panel we have randomly permuted the  true predictor coefficients
for each realization, thereby causing the monotonicity to be violated
(on average), but keeping the same signal-to-noise ratio.  Not surprisingly, the
 ordered lasso does better when the true coefficients are monotone, while the reverse is true for the lasso. However, we also see that in an absolute sense one can achieve a much lower MSE in the monotone setting
of  the left panel.
\subsection{Performance on Los Angeles ozone data}
\label{sec:ozone}
These data are available at \url{http://statweb.stanford.edu/~tibs/ElemStatLearn/data.html}.
They represent the level of atmospheric ozone concentration from
eight daily meteorological measurements made in the Los Angeles basin
  for 330 days in 1976. 
 The response variable
is the log of the daily maximum of the hourly-averaged ozone
concentrations in Upland, California.
We divided the data into  training and validation
sets of approximately the same size, and considered models with a maximum time-lag of 20 days.

\begin{figure}
\centering
\includegraphics[scale = 0.45]{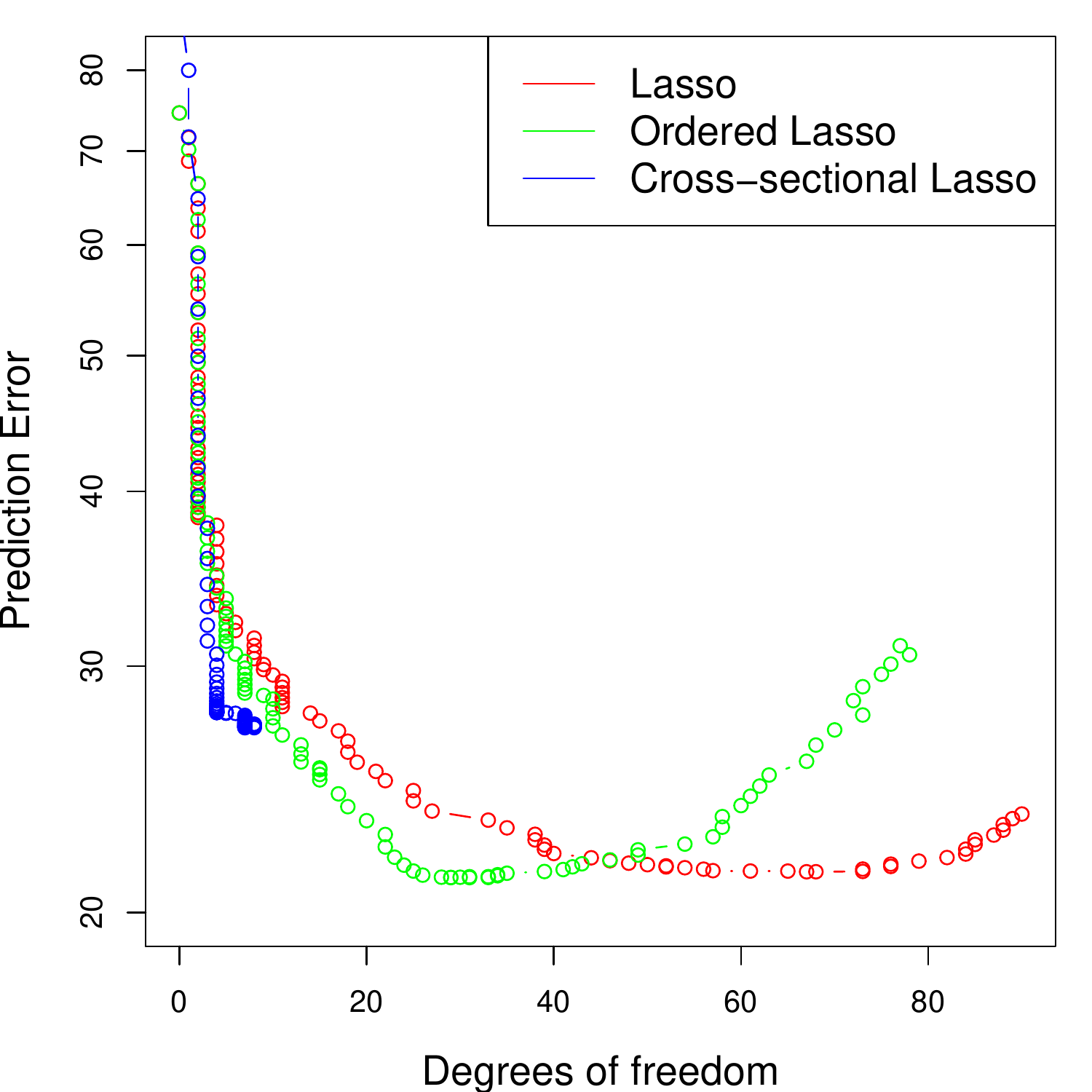}
\caption{\textit{Ozone data: prediction  error curves.  The cross-sectional lasso
(blue)  predicts from measurements on the same day, the lasso(red) predicts from
measurements on the same day and previous 19 days, and the ordered lasso (green)
adds the monotonicity
constraint to the lasso.}}
\label{fig:ozone}
\end{figure}

Figure \ref{fig:ozone} shows the prediction error curves over the validation set,
for  the ``cross-sectional'' lasso (predicting from measurements on the same day), the
lasso
(predicting from measurements on the same day and the previous 19 days),
and the ordered lasso, which adds the monotonicity constraint to the lasso.
We see that the  ordered lasso and the lasso applied to time-lagged features achieve lower errors
than the ``cross-sectional'' lasso. In addition, the  ordered lasso achieves the minimum with
fewer degrees of freedom(as defined in Section \ref{sec:df}).

\begin{figure}[h]
\centering
\includegraphics[width = 5in, height = 2.5in]{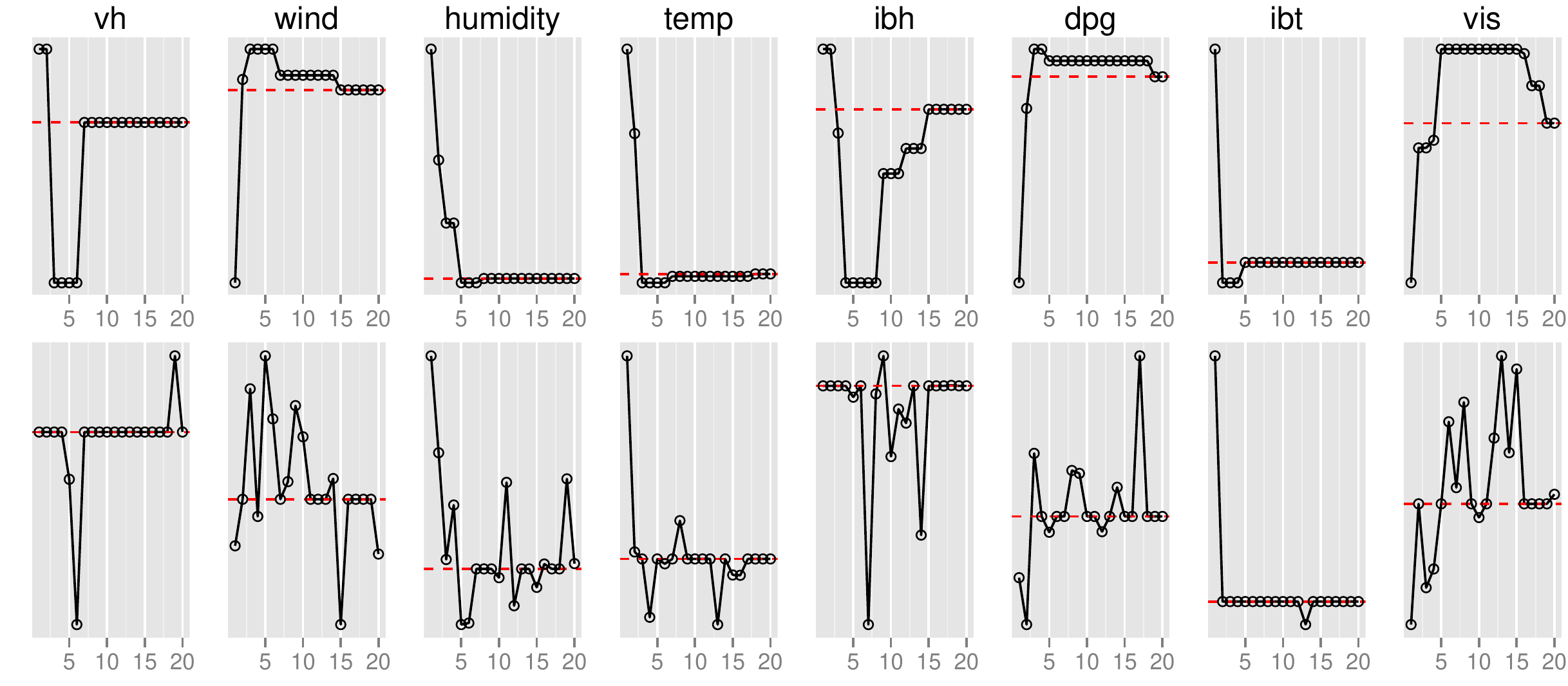}
\caption{\textit{Ozone data: estimated coefficients from the ordered lasso (top)
and the lasso (bottom) versus time-lag. For reference,  a dashed red horizontal line is drawn at
zero.}}
\label{fig:ozonecoef}
\end{figure}

Figure \ref{fig:ozonecoef} shows the
estimated coefficients from the ordered lasso (top) and the  lasso
(bottom). The  ordered lasso yields simpler and  more interpretable solutions. For each predictor,
the ordered lasso also 
determines the most suitable estimate of  the time-lag interval, beyond  which the estimated coefficients are zero.
For example, the estimated coefficients of the predictor ``wind"  are zero beyond a time-lag of 14 days from the current time $t,$ whereas  the estimated coefficients ``humidity" are zero beyond a  time-lag of 7 days from the current time $t$.    
\vspace*{-0.4cm}
\subsection{Auto-regressive time series applied to sunspot data and simulated data}
In an auto-regressive time series model, one predicts each value $y_t$ from
the   values $y_{t-1}, y_{t-2} \ldots y_{t-k}$ for some maximum lag, or
 ``order'' $k$. This fits into the time-lagged regression framework, where the regressors are simply the
time series itself at previous time points.
Our proposal for  monotone constraints
in the AR model seems to be novel. \citeasnoun{Nardi2011528} studied the application
of the standard lasso to the
AR model  and  
derived  its  asymptotic properties. 
\citeasnoun{JTSA:JTSA12027}
suggested  a Bayesian approach  to the lasso based on  the partial autocorrelation
representation of AR models. In the following example, we compare coefficient estimates
and order estimates among the ordered lasso, the lasso, and the standard AR fit.

\begin{figure}[htp]
\centering
\includegraphics[width = 10cm,height = 6cm]{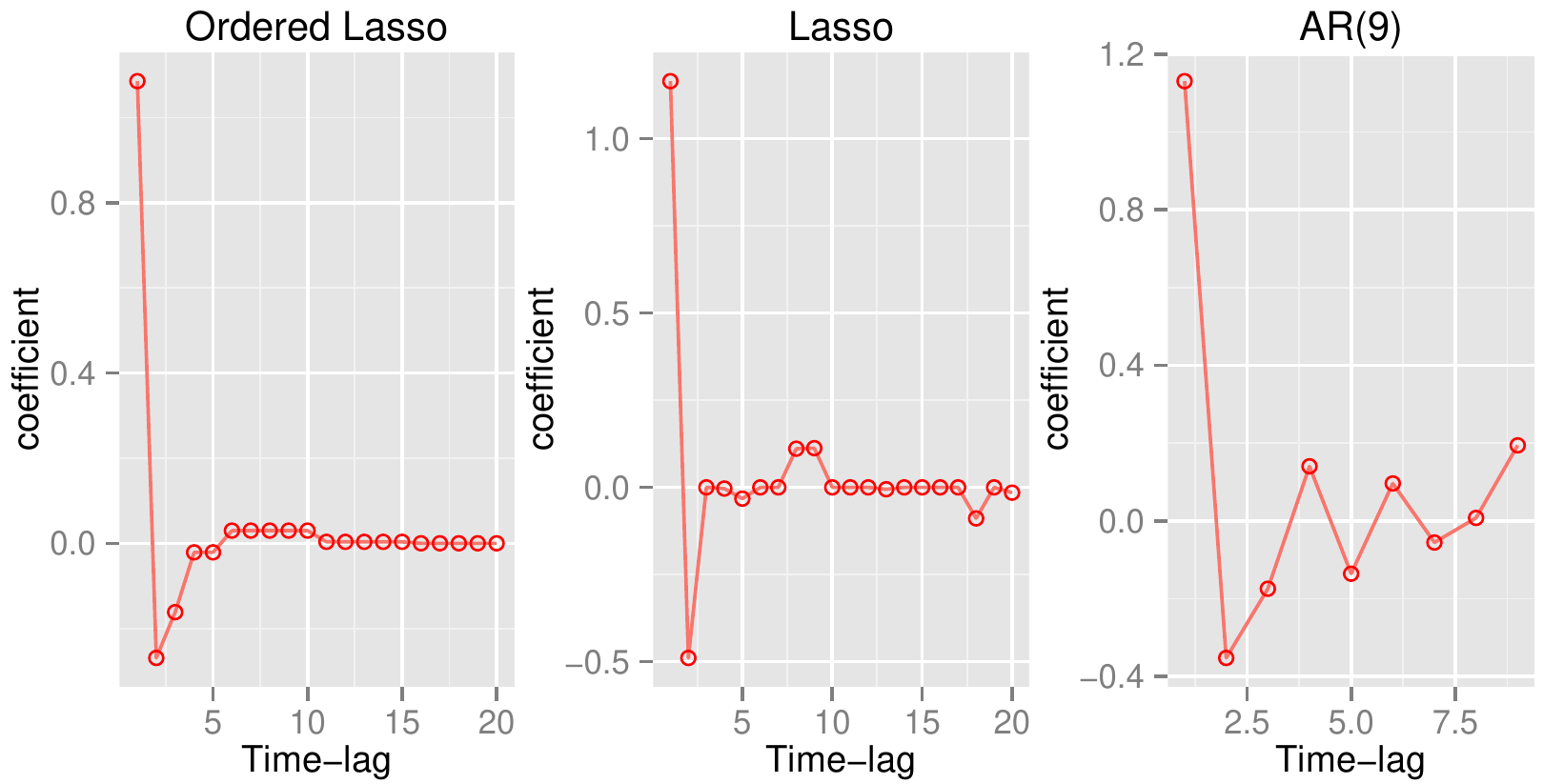}
\caption{\it Sunspot data: estimated coefficients of the ordered lasso,the lasso and  the standard AR fit. }
\label{fig:sunspot}
\end{figure}

The data for this example is available in the  R package as {\tt sunspot.year}. The data contains
289 measurements and they represent yearly numbers of sunspots from  1700 to 1988.
Figure \ref{fig:sunspot} shows the results of the auto-regressive model fit to the yearly sunspot data. We separated the series into training and validation series of about equal size.
The  standard AR fit (right panel)  chose an order of 9 using least squares and
the Akaike information criterion (AIC).
The ordered lasso (with $\lambda$ chosen by two-fold cross validation) suggests an order of 10 (out of a maximum of 20)
and  gives a well-behaved sequence of coefficients.
The regular lasso (middle panel) --- with no monotonicity constraints--- gives a less clear picture.
All three estimates had about the same error on the validation set.
\setlength{\extrarowheight}{0.26cm}
\begin{table}
\centering
\begin{small}
\begin{tabular}{l|rrrrrrrrrr}\Xhline{4\arrayrulewidth}
\diag{.2em}{3cm}{Method}{Est. lag}&1&2&3&4&5&6&7&8&9&10\\\hline
AR/AIC& 0&0&69&14&6&2&3&3&0&3\\\hline
Ordered Lasso    & 0&0&66&14&3&5&4&1&3&4\\\Xhline{4\arrayrulewidth}
\end{tabular}
\end{small}
\caption{\it Estimates  of AR lag from  the ordered lasso, and AR model using least
squares and AIC from 100 simulations. The data was generated as $y_i = \sum_{k =
1}^3y_{i-k}\beta_k+\sigma\cdot
Z_{i}$
where $\sigma = 4, y_{i},Z_i\sim N(0,1)$, and $\bbeta = \{0.35,0.25, 0.25\}.$ Each
entry represents the number of times that a specific lag was estimated in 100 simulations.
 }
\label{tab:order}
\end{table}

Table \ref{tab:order} shows the results 
of an experiment comparing the ordered lasso  to the standard AR fitting using AIC
 from
100 simulations.
The goal was to estimate the 
 lag of the time series (number of non-zero coefficients) , as in the previous figure. The true series was of length 1000, with an actual  lag of 3,
and the maximum lag considered was 10. The data was divided into training and validation
series of approximately the same size. The ordered lasso used the second half of the series to estimate the best value of $\lambda$ and   estimate the order of the series. The results show that the ordered lasso has similar performance to AR/AIC for this task.
\section{Degrees of freedom}
\label{sec:df}
Given a fit vector $\hat\by$ for estimation from a vector $\by \sim N(\bmu, \bI\cdot\sigma^2)$,
  the degrees of freedom of the fit can be defined as
 \begin{eqnarray}
         {\rm df}(\hat\by)=\frac{1}{\sigma^2}\sum_{i=1}^N{\rm Cov}(y_i,\hat y_i)
         \label{eqn:cov}
 \end{eqnarray}
 \cite{Ef86}. 
 This applies even if $\by$ is an adaptively chosen estimate.
 \citeasnoun{lassodf} show that for the lasso, the number of non-zero ``plateaus'' (coefficients) in the solution is 
 an unbiased estimate of the degrees of freedom. \citeasnoun{TT2011}
 give analogous estimates for generalized penalties.
 For  near-isotonic regression
 described in Section \ref{NearIso},
 letting $\hat k$ denote  the number of nonzero ``plateaus" in the solution,
 \citeasnoun{THT2011} show that
 \begin{eqnarray}
 {\rm E}(\hat k)={\rm df}(\hat\by)
 \label{eq:dford}
 \end{eqnarray}
 For the ordered lasso, this can be applied directly in the orthogonal
 design case to yield (\ref{eq:dford}).
 For the general $\bX$, 
we conjecture that the same result holds, and can be established by 
 studying the properties of   projection onto the convex constraint set 
 (as detailed in \citeasnoun{TT2011}).
\section{Logistic regression model}
\label{sec: logistic}
Here we show how to generalize the ordered lasso to logistic regression. Assume that
we observe $(\bx_i, y_i), i=1,2,\ldots ,N$ with $\bx_i=(x_{i1}, \ldots ,x_{ip})$
and $y_i=0$ or $1$. The log-likelihood function is
\begin{eqnarray*}
        l(\bbeta) = \sum_{i=1}^{N}(y_i(\beta_0+\bx_i^T\bbeta )- \log(1+e^{\beta_0+\bx_i^T\bbeta})).
\end{eqnarray*}
With the ordered lasso, we write each $\beta_j=\beta_j^+ -\beta_j^-$ with $\beta_j^+, \beta_j^-\geq
0$,    and solve 
\begin{eqnarray}
            {\text{maximize}} \{l(\bbeta^+-\bbeta^-)-  \lambda(\sum_{j
                = 1}^p(\beta^+_j+\beta_j^-)\},
\label{eqn:logist}
\label{eqn:logist}
\end{eqnarray}
subject to   $\beta^+_1\geq \cdots \geq \beta^+_p \geq 0$ and $\beta^-_1 \geq \cdots \geq \beta^-_p\geq 0$. 
We write our data in matrix form and use  the iteratively reweighted least squares method (IRLS) to   solve  (\ref{eqn:logist}), i.e.,
at each iteration, we solve 
\begin{eqnarray}
 {\text{minimize}} &\Bigl\{\frac{1}{2}(\bz-\beta_0- \bX(\bbeta^{+}-\bbeta^-))^T\bW(\bz
- \beta_0- \bX(\bbeta^{+}-\bbeta^-)) \nonumber \\ & + \lambda \sum_{i = 1}^p(\beta^+_i+\beta_i^-)\Bigr\},
\label{eqn:weigthedleastsqaure}
\end{eqnarray}
 subject to $\beta^+_1\geq \beta^+_2 \geq \cdots \geq \beta^+_p\geq 0$ and $\beta^-_1\geq \beta^-_2 \geq \cdots \geq \beta^-_p\geq 0$, where $\bz =\beta_{0}^{\rm old}+ \bX(\bobeta^+-\bobeta^-) + \bW^{-1}(\by-\bp)$, $\bp$ is a vector with $\bp_i  = \frac{\exp(\beta^{\rm{old}}_0+\bx_i^T(\bobeta^+-\bobeta^-))}{1+\exp(\beta_0^{\rm
old}+\bx_i^T(\bobeta^+-\bobeta^-))}$, and $\bW$ is a diagonal matrix with $\bW_{ii} = \bp_i(1-\bp_i)$.
We apply the ordered
lasso (Algorithm
\ref{alg: orderedlassoalgorithm})  to solve (\ref{eqn:weigthedleastsqaure})
with  modified updates:  
\begin{eqnarray*}
&&\textstyle\beta_0 \leftarrow\beta_0-\gamma {\bf 1}^T\bW(\beta_0+\bX\bbeta-\bz)
, \\&&\bbeta\leftarrow{\bprox}_{\gamma\lambda}(\bbeta-\gamma\bX^T\bW(\beta_0+\bX\bbeta-\bz)).
\end{eqnarray*}
Applying the ordered lasso to the logistic regression model with  time-lagged features,  we approximate the log-likelihood function as  in (\ref{eqn:weigthedleastsqaure})  and use Algorithm \ref{alg: statictimelaglasso} or Algorithm \ref{timeLagorderedlasso} to solve the weighted least squares minimization subproblem.  Similar extensions can
 be made to other generalized linear models. 
\section{Discussion}
\label{sec:discussion}
In this paper, we have proposed an order-constrained version of the lasso.
This procedure has natural applications
to the static  and rolling prediction problems, based on time-lagged variables.
It can be applied to any dynamic prediction problem, including  financial time series and prediction
of dynamic patient outcomes based on  clinical measurements.
For the future work, we could generalize our framework to higher dimensional notions
of monotonicity, which could be useful for spatial data.
An R package that implements the algorithms will be made available on the CRAN website.
\section{Acknowledgement}
We thank Stephen Boyd for his helpful suggestions.
\bibliographystyle{agsm} 
\bibliography{tibs}
\end{document}